# Performance Evaluation of the Labelled OBS Architecture

Thomas Legrand[(1)], Bernard Cousin[(1)] and Nicolas Brochier[(2)]
[(1)]University of Rennes 1, IRISA,
Campus universitaire de Beaulieu, F-35042, Rennes, France
[(2)]France Telecom, Orange Labs,
2 avenue Pierre-Marzin, F-22307, Lannion cedex, France

*Abstract-* **A comparison of three different Optical Burst Switching (OBS) architectures is made, in terms of performance criteria, control and hardware complexity, fairness, resource utilization, and burst loss probability. Regarding burst losses, we distinguish the losses due to burst contentions from those due to contentions of Burst Control Packets (BCP). The simulation results show that as a counterpart of an its additional hardware complexity, the labelled OBS (L-OBS) is an efficient OBS architecture compared to a Conventional OBS (C-OBS) as well as in comparison with Offset Time-Emulated OBS (E-OBS).**

I. INTRODUCTION

The evolution of optical transport networks is driven by continuously increasing traffic demand and Internet applications. Optical Burst Switching (OBS) architectures are considered to be promising solutions for coping with these trends [1-2].

At the ingress node of OBS networks, the client data with the same egress node are aggregated and assembled into a large data unit called a 'burst'. When the burst is assembled, the edge node builds a Burst Control Packet (BCP) that contains control information for routing operations along the transmission path. The BCP is then transmitted on a specific control wavelength and is sent prior to the associated burst with an appropriate Offset Time (OT). According to the management of OT, we define Conventional OBS (C-OBS) and Offset Time-Emulated OBS (E-OBS) [1]-[3]. C-OBS uses a variable OT whereas E-OBS uses a fixed OT. When a BCP arrives at a core node, it's processed electronically by the control unit (CU) of the node which configures the switching matrix of the node before the burst arrival. Thus, the burst can be transmitted optically toward the next node of the forwarding path. At the destination node, the burst is converted in the electronic domain. The burst is then disassembled to recover the client data.

More recently, we proposed an OBS architecture called the Labelled OBS (L-OBS) [4-5]. Contrary to the two previous OBS architectures, control information is carried by a label, and the label and its burst are sent successively and on the same wavelength. A particular gain of the L-OBS architecture is to reduce the control complexity without increasing the burst loss probability.

Whatever the OBS architecture, a well-known drawback is burst losses due to burst contentions. A burst contention occurs at core node when at least two bursts require the same output port and the same wavelength at the same time. This problem has been largely studied in the literature. Hence, different mechanisms of burst contention resolution and scheduling algorithms have been developed in order to minimize these burst losses [6-10]. However, BCP losses also bring about burst losses. Thus, congestions of control units of nodes and BCP contentions increase the burst loss probability. Whereas the problem of congestion of the control unit is discussed in [11], to the best of our knowledge BCP contentions have never been studied.

In this work, we compare C-OBS, E-OBS, and L-OBS concepts in terms of complexity, burst loss probability, fairness, and resource utilization. We also distinguish losses due to burst contentions from losses due to BCP contentions.

The rest of this paper is organized as follows. Section II describes the operating principle of C-OBS, E-OBS, and L-OBS networks, and discusses the control and hardware complexity. Section III presents the simulation environment and describes the reservation protocols and scheduling algorithms used by the simulations. Section IV compares C-OBS, E-OBS, and L-OBS networks in terms of burst loss probability, fairness, and resource utilization. Section V concludes the paper.

II. OPERATING PRINCIPLE

We consider three different OBS architectures varying in the management of their signalling. The first one is the C-OBS architecture. It uses an out-of-band signalling with variable OT between the bursts and their BCP. The second one is the E-OBS architecture. It employs an out-of-band signalling with fixed OT. The last one is the L-OBS architecture, using an in-band signalling. The following section describes in more detail these architectures.

*A. Conventional OBS (C-OBS)*

In C-OBS networks [1-2], the OT is set once at the ingress nodes. It must provide enough time for processing the control information of the BCP and for configuring the switching matrix at each core node along the transmission path before the burst arrival (Fig. 1(a)). Thus, the value of the OT is based on the path length. After having sent the BCP, when the OT expires, the ingress node releases the burst that crosses the configured nodes all along the transmission path. Thus, the burst crosses the network while staying in the optical domain and without delay.

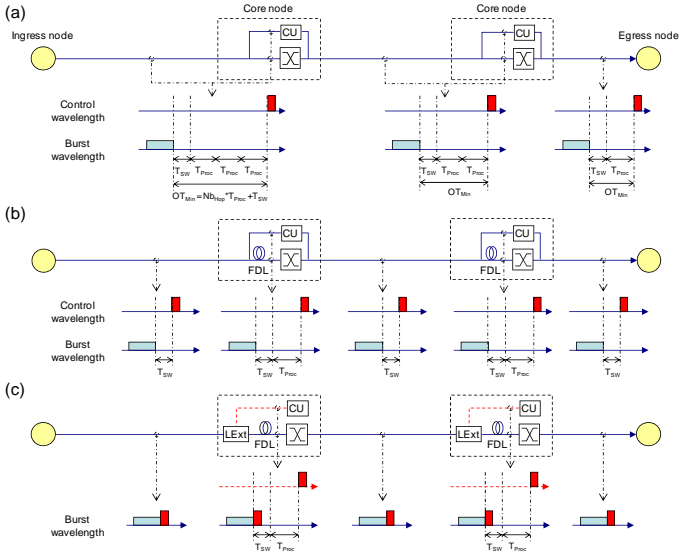

Fig. 1: Example of a burst transmission in C-OBS (a), E-OBS (b) and L-OBS (c). $T_{SW}$ is the Switching Time, $T_{Proc}$ is the Processing Time, $Nb_{Hop}$ is the number of hops on the forwarding path and $OT_{Min}$ is the Minimum OT.

As shown in Fig. 1(a), the time advance between the BCP and its burst decreases with the BCP processing time in every node along the burst path. Therefore, the ingress node must define a minimum OT ($OT_{Min}$) so as to guarantee that bursts don't overtake their BCP. This $OT_{Min}$ is a function of processing time ($T_{Proc}$), number of hops ($Nb_{Hop}$) in the forwarding path, and switching time ($T_{SW}$).

$$OT_{Min} = T_{SW} + \sum_{i=1}^{Nb_{Hop}} T_{Proc} \qquad (1)$$

Moreover, the value of the OT must be updated at each path node in order to maintain an accurate knowledge of the time between the burst and its BCP. This management of OT increases the control complexity.

From the hardware complexity point of view, the C-OBS node is the simplest of the three OBS nodes. It may consist solely of a CU and a switching matrix (Fig. 1(a)). According to a scheduling algorithm (see section III.A), CU reserves the resources for the burst transmission request carried by the incoming BCP and it configures the switching matrix at the burst arrival. In this OBS network, the time spent in the CU for processing control information can fluctuate. But, a fixed time processing ($T_{Proc}$) simplifies OT management and control complexity. In return, it introduces a possible BCP contention on the output control wavelengths. In the following, we consider fixed time processing because it's required by the two other OBS networks.

For C-OBS, the end-to-end delay transmission ($T_{End-to-end}$) can be calculated as the summation of burst assembly time ($T_{Ass}$), $OT_{Min}$ and propagation delay along the path ($T_{Prop}$). So, it's defined as

$$\begin{aligned} T_{End-to-end} &= T_{Ass} + T_{Prop} + OT_{Min} \\ &= T_{Ass} + T_{Prop} + T_{SW} + \sum_{i=1}^{Nb_{Hop}} T_{Proc} \end{aligned} \qquad (2)$$

*B. Offset Time Emulated OBS (E-OBS)*

In E-OBS architectures [1]-[3], a BCP is sent prior to its burst with an OT at least equal to the time necessary for configuring the switching matrix ($T_{SW}$). When this OT expires, the burst is released by the ingress node. At each input port of the core nodes, a Fibre Delay Line (FDL) is inserted on the burst path so as to delay it during the processing of its BCP. As a result, this FDL compensating the processing time of the BCP enables to keep a fixed OT all along the forwarding path (Fig. 1(b)).

The E-OBS node is more complex than the C-OBS node because of the introduction of one FDL for each node input. In addition, a strict control of the time spent by BCPs in the CU is necessary for dimensioning the input FDLs and to keep a fixed OT from link to link inside the network. The inconvenience of this fixed processing time is the possible contention between BCPs for access to the control wavelength.

In comparison with the C-OBS network, the E-OBS network using a fixed OT reduces the control complexity both at BCP processing and OT management. Indeed, no OT management is necessary at the core nodes and at the ingress node, the setup of the OT value is independent of the forwarding path. A particular advantage of the fixed OT is that the utilization of source routing or hop-by-hop routing is easy to implement in E-OBS network [3], whereas in C-OBS networks, the hop-by-hop routing involves an additional control mechanism [12-13] in order to guarantee that the burst doesn't overtake its BCP. Moreover, in the C-OBS network, the update of the network topology or even the update of a core node will lead to the update of $OT_{Min}$ values at all edge nodes of the network. In the E-OBS network, this effect is resolved by the use of a fixed OT.

Finally, E-OBS end-to-end delay is equal to C-OBS end-to-end delay (Fig. 1(a) and (b)). So, it's defined by (2).

*C. Labelled OBS (L-OBS)*

In an L-OBS network [4-5], the burst is composed of a data section and a header section called a 'label'. This label carries control information required for the routing operations along the forwarding path. In this network, the client data and the control information are sent together on the same wavelength.

It's in-band signalling. Therefore, the optical label must be read and processed electronically at each node of the forwarding path while the burst is delayed by an input FDL (Fig. 1(c)).

From the hardware complexity point of view, the L-OBS node is the most complex of the three types of OBS node. In addition to input FDLs, a Label Extractor (LExt) is necessary. Located before the input FDLs, this function extracts the control information carried by the optical label and converts it in the electronic domain (Fig. 3). Then, the electronic label is sent to the CU which processes it and configures the switch matrix according to the result of a scheduling algorithm. In an L-OBS network, the time spent between the arrival of a label at the CU and the end of the configuration of the switching matrix must be the same from one label to another in order to properly define the length of the input FDLs.

The end-to-end delay in L-OBS is the longest. Indeed, unlike C-OBS and E-OBS networks, the processing time and the switching time have to be compensated at each hop (Fig. 1(c)). So, the end-to-end delay is expressed as

$$T_{End-to-end} = T_{Ass} + T_{Prop} + \sum_{l=1}^{Nb_{Hop}} (T_{Proc} + T_{SW}) \qquad (3)$$

Nevertheless, in the usual optical switching matrix, the switching time is below one µs (or a few ns for the fastest optical switching technologies [14]); this is very small in comparison with the propagation delay (1 ms for a 200 km link) and it remains short compared to the processing time (some µs [4-5]). Consequently, the end-to-end delay of L-OBS is close to those of C-OBS and E-OBS networks.

In comparison with C-OBS and E-OBS, the L-OBS network reduces the control complexity. For C-OBS and E-OBS networks, in a meshed topology, synchronization of the BCP with the burst can be very difficult to ensure. Moreover, the burst time location is calculated from the BCP arrival time and its control information (OT and burst duration). So, a stringent synchronization of the clock of different nodes inside the network is required in order to accurately define the burst time location. Furthermore, for C-OBS and E-OBS, in case of a failure (hardware or software) on the forwarding path of BCPs, loss of synchronization would be unavoidable and lead to network operation instability. Finally, for C-OBS and E-OBS, a failure on the control wavelength of a link (e.g., a failure either of the transmitter or receiver of the control wavelength) results in the failure on the totality of the link. These issues can be addressed by using labels, as in L-OBS networks. Indeed, all computations are performed locally at the node so the problems of synchronization in C-OBS and E-OBS networks are resolved. Moreover, any wavelength may transmit the bursts and their control information carried by the label, thus a failure on one wavelength (e.g., a failure either of the transmitter or receiver of the wavelength) leads only to the loss of transmission capacity of this wavelength. Finally, similarly to the E-OBS network (see Section II.B), the L-OBS architecture resolves the problem of flexibility in terms of routing and update of network topology encountered in a C-OBS network.

III. SIMULATION ENVIRONMENT

In our simulation scenario, we consider two network topologies. The NSFNET topology is a real topology often used in similar studies (e.g., in [3]). It is composed of 14 nodes and 21 bidirectional links. The second topology is a regular topology. We selected this topology in order to limit the impact of the unbalanced load on the network performance. It's a 36-node torus with 72 bidirectional links. We suppose that all network links have the same number of wavelengths W=32. For C-OBS and E-OBS, one wavelength of each link is a Control Wavelength (CW=1) which is dedicated to forwarding BCPs. The transmission bit rate of bursts and control packets (BCP and label) is 10 Gbps. In these topologies, we apply shortest path routing.

The traffic is uniformly distributed between the nodes. Each node sends the same amount of traffic to any other node. The offered traffic by one edge node is expressed in Erlangs and is normalized to the transmission capacity of a link. In other words, one Erlangs corresponds to the amount of traffic that occupies one entire link. The bursts are generated according to an exponentially distributed arrival process and have an exponentially distributed length. The mean duration of bursts is 100 µs (i.e., 1 Mbit at 10 Gbps). Concerning control packet generation (BCP or label), we consider a fixed length of 100 bits (i.e., 10 ns at 10 Gbps).

We assume that each node is an edge and a core node. The core node can resolve burst contentions in the spectral domain. The switching and processing times are fixed respectively at 1 µs and 10 µs.

For most of the results presented in this paper, the parameters used are as explained above. Only the results of Fig. 4 are obtained with different parameters: these are discussed in Section IV.A.

A. *Reservation protocol and scheduling algorithm*

In our simulation, we use either Just Enough Time (JET) or Horizon as reservation protocols [1-2]-[15-16]. Similarly, we use either Last Available Unscheduled Channel (LAUC) or Last Available Unscheduled Channel with Void Filling (LAUC-VF) as scheduling algorithms [6-8].

Both JET and Horizon use a delayed reservation technique. So, their resource reservations begin at the arrival time of the incoming bursts, and the duration of their reservations is equal to the length of bursts. The difference between these algorithms is that for each wavelength, Horizon memorizes the earliest time just after the end of the last reservation while JET memorizes all the reservation periods. So, JET enables the filling of gaps between the previous reservation periods. In counterpart, it's more complex than Horizon.

LAUC is a delay-oriented scheduling algorithm. Therefore, when a new burst reservation request arrives, it chooses the wavelength providing the shortest delay on the burst transmission. When several wavelengths are possible, it selects the one that minimizes the gap generated between the previous reservation and the new burst reservation so as to increase the channel utilization. Conversely, if all wavelengths are unavailable then the incoming burst is lost. LAUC-VF is close to LAUC. The only difference is that LAUC-VF can fill the gaps that occur between the reservations. But, it's also the most complex of the two algorithms.

In C-OBS, due to OT variation, the arrival order of bursts isn't always the same as the arrival order of their BCPs. Consequently, gap may occur between two reservations. In such case, LAUC-VF and JET appear as relevant algorithms for C-OBS network. On the other hand, in E-OBS and L-OBS networks, the time separating the burst from its control packet (label or BCP) is the same for all the bursts. So, the arrival order of bursts is the same as the arrival order of their control packets. Consequently, gap can't be created between the burst reservations and thus, the void filling concept is not necessary.

As a result, for C-OBS network, we perform JET and LAUC-VF whereas for E-OBS and L-OBS networks, we carry out Horizon and LAUC.

## IV. PERFORMANCE COMPARISON

In this section, we compare the performance of C-OBS, E-OBS and L-OBS networks in terms of burst loss probability, fairness and network resource utilization.

### A. Burst loss probability

In Fig. 2, we present the burst loss probability as function of offered load for C-OBS, E-OBS and L-OBS networks. The continuous lines are the simulation results obtained on the NSFNET topology whereas the dashed lines represent the results on the Torus topology. Whatever the offered load and whatever the topology, the L-OBS network has better performance than the E-OBS which is itself more efficient than the C-OBS. We recall that only C-OBS network performs LAUC-VF and JET which can fill the voids. Nevertheless, the C-OBS network has the worst performance.

We present in the Fig. 3 the burst loss probability due to either burst contention (Fig. 3(a)) or BCP contention (Fig. 3(b)). We can see that the burst losses due to BCP contentions aren't negligible for C-OBS and E-OBS. As a matter of fact, BCP contentions are the main source of burst losses up to an offered load of 0.4. So, under this limit the burst loss probability observed in Fig. 2 for C-OBS and E-OBS is the result of BCP contentions. Moreover, we observe that this burst loss probability is the same for C-OBS and E-OBS. As discussed in sections II.A and II.B, it's the use of a fixed processing time that leads to burst losses due to BCP contentions in C-OBS and E-OBS networks (see Fig. 3(b)).

For an offered load greater than 0.4, the burst contentions become predominant for all the OBS architectures (Fig. 3(a)). However, the L-OBS is the most effective whatever the offered load and the network topology. On the other hand, the C-OBS is the least efficient. A fact justifying the poor performance of C-OBS is that the burst loss probability in C-OBS network is OT-dependent because of OT variation. This OT dependency is already well-known in the literature [17] and it leads the development of Quality of Service (QoS) mechanism [18]. In fact, a burst with a long OT has more chance to reserve an output wavelength than a burst with a short OT. As a consequence, the burst loss probability increases when the burst approaches its destination. These burst losses close to their destination lead to the waste of transmission resource reserved (see section IV.C) and the increasing of burst loss probability seen in the Fig. 3(a).

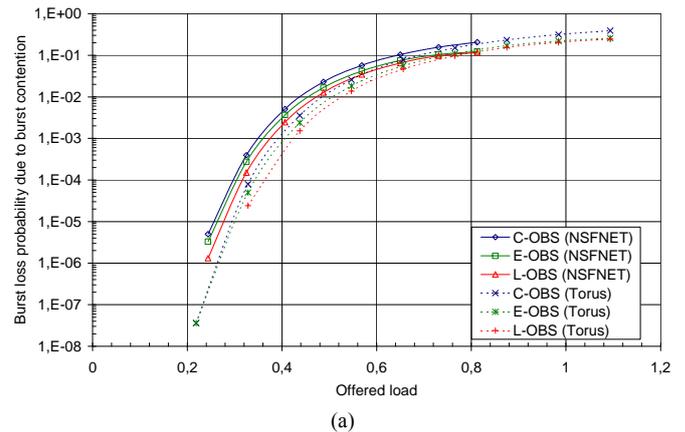

(a)

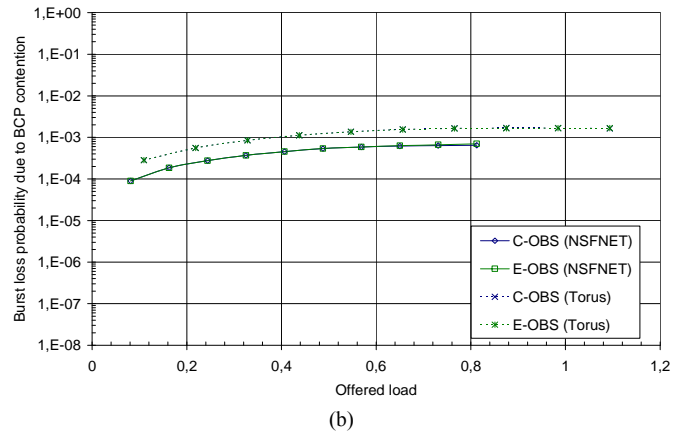

(b)

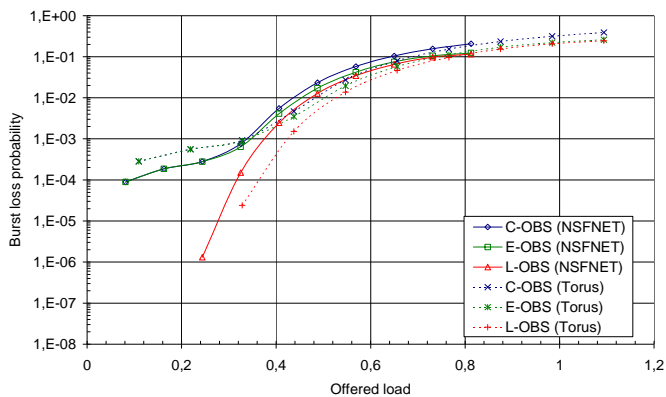

Fig.2: Burst loss probability as function of offered load. C-OBS uses JET and LAUC-VF whereas E-OBS and L-OBS use Horizon and LAUC.

Fig. 3: Burst loss probability due to either burst contention (a) or BCP contention (b) as a function of offered load for C-OBS, E-OBS and L-OBS.

In the case of C-OBS and E-OBS networks, at least one control wavelength is assigned to the forwarding of BCPs whereas the L-OBS network uses all wavelengths of the link for sending bursts. Consequently, L-OBS has a better utilization rate of network resources (see section IV.C) and it enhances the statistical wavelength multiplexing between the bursts. These effects justify the best performance of L-OBS networks regarding the burst loss probability due to burst contentions regardless of the offered load (Fig. 3(a)).

In Fig. 4, we compare the impact of various parameters on the BCP contentions. It shows the burst loss probability due to BCP contention as a function of offered load with different simulation parameters. Here, only the results obtained with the C-OBS network are presented. In fact, as we showed in Fig. 3(b), the results obtained with the E-OBS network would be close. For readability, we only present results on the NSFNET topology. In Fig. 4, we modify the number of wavelengths per link (W=16 or W=32), the average burst length ($L_B$=1 Mbit, i.e., 100 µs at 10 Gbps, or $L_B$=5 Mbit, i.e., 500 µs at 10 Gbps) and the transmission bit rate of BCPs ($α_{BCP}$=10 Gbps, i.e., 10 ns for 100 bits, or $α_{BCP}$=622 Mbps, i.e., 160 ns for 100 bits).

To exploit these results, we consider the results with the previous parameters (W=32, $L_B$=1 Mbit and $α_{BCP}$=10 Gbps) as the baseline. Compared with this baseline, we show that the burst loss probability is proportional to the number of wavelengths per link and on the contrary, is inversely proportional to both the length of bursts and the transmission bit rate of BCPs. As a result, the use of a low bit rate technology for BCP transmission in order to reduce the cost brings about a worsening of network performance. Similarly, the increasing of the number of wavelengths with the aim to raise the transport capacity degrades the network performance in terms of BCP contentions.

B. *Fairness*

In this section, we focus on the fairness of bursts in terms of burst loss probability. We have already seen that in the C-OBS network, the burst losses depend on the OT (see section IV.A). But, we didn't evaluate the fairness in E-OBS and L-OBS. To this end, Fig. 5 assesses the burst loss probability with respect to the number of remaining hops to reach the destination for C-OBS, E-OBS, and L-OBS architectures. In these results, we consider the two topologies with the same set of parameters presented in Section III. For each result, the offered load is 0.65.

In Fig. 5, we can see the poor fairness of a C-OBS network. Indeed, the bursts having the greatest number of remaining hops have more chance to reserve an output wavelength. Therefore, the bursts starting their trip may undergo much lower losses than the bursts having a last hop to reach their destination.

As discussed in section III.A, in L-OBS and E-OBS, all the bursts arriving at the same time start their reservation of resources at the same time. Thus, they have the same chance to reserve an available output wavelength. The results of Fig. 5 confirm this behaviour and prove that E-OBS and L-OBS networks address the problem of fairness of the C-OBS network. The slight variation observed for the NSFNET topology is due to unbalanced load. This assumption is validated by the results obtained for the regular torus topology which present a steadier burst loss probability.

The difference of performance between the L-OBS and E-OBS architectures is the result of BCP contentions and the reservation of a specific wavelength for the BCP forwarding.

C. *Resource utilization*

Network resource utilization is an important metric from the operational point of view. To compare the performances of C-OBS, E-OBS, and L-OBS in resource utilization, we present this metric as a function of offered load (Fig. 6). Here, we define the network resource utilization as the network resources reserved by the bursts successfully delivered at their destination divided by the total network transmission capacity. The simulation parameters are identical to those used in section III.

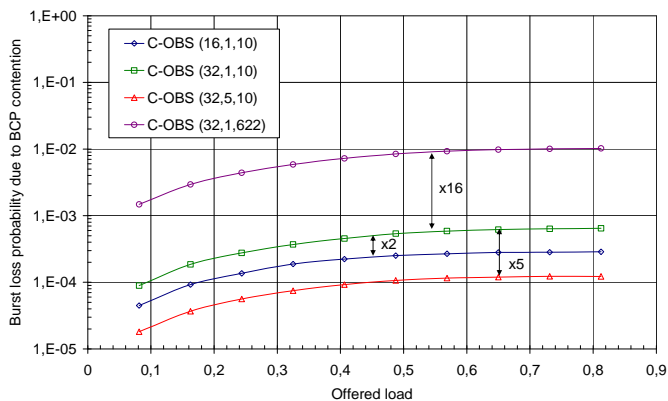

Fig. 4: Impact of number of wavelengths, burst length and bit rate of BCP on the burst loss probability due to BCP contention in C-OBS.

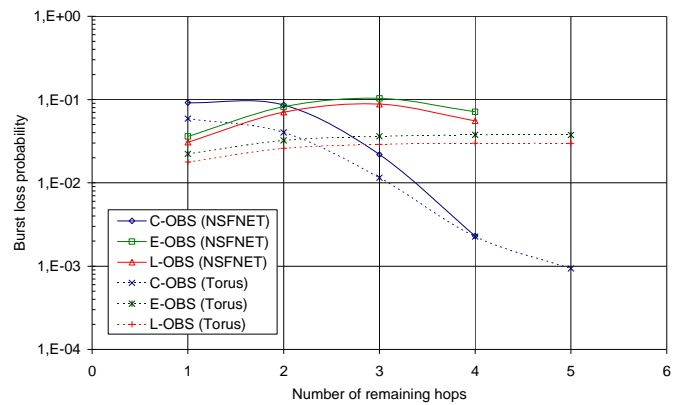

Fig. 5: Burst loss probability as a function of number of remaining hops for C-OBS, E-OBS and L-OBS. The offered load is 0.65.

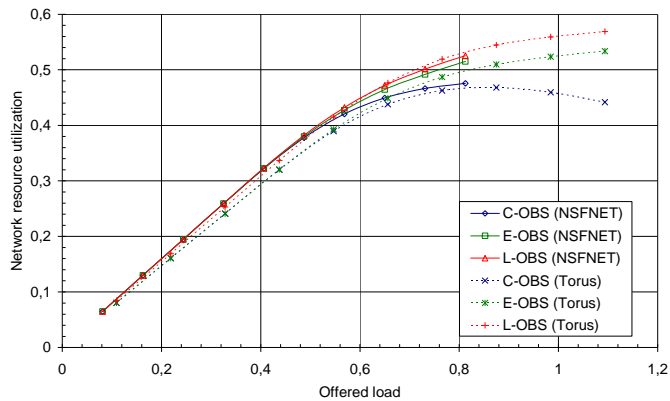

Fig. 6: Comparison of network resource utilization

According to Fig. 6, the L-OBS network surpasses the two other OBS architectures and the C-OBS is the least efficient. Once more, the poor performance of C-OBS is explained by the OT variation. In fact, the C-OBS architecture wastes network resources in order to transmit bursts which will be dropped when they are close to their destination. Compared to an L-OBS network, an E-OBS network performs slightly worse. This is due to the use of a control wavelength and to BCP contentions, both of which increase the burst loss probability of the E-OBS network (Figs. 2 and 3). Moreover, in the regular torus topology, the difference of performance between the OBS networks is increased because the saturation effects due to the unbalanced load in the NSFNET topology are reduced.

V. CONCLUSION

In this paper, the L-OBS network is the best OBS architecture in terms of burst loss probability and network resource utilization. In addition, we showed that the access unfairness to network resource of C-OBS (for bursts having different OT) disappears in E-OBS and L-OBS networks.

When a fixed processing time is considered, BCP contentions have an impact on the burst loss. Furthermore, BCP contentions are proportional to the number of wavelengths and inversely proportional to both the length of bursts and the transmission bit rate of BCPs.

In summary, the L-OBS architecture resolves the problems of control complexity and OT management which are inherent in C-OBS and E-OBS networks, in compensation for some additional hardware complexity. In addition, the L-OBS is the most efficient OBS architecture in terms of burst loss probability and resource utilization.